\documentclass[letterpaper, english, aps, prl, reprint, twocolumn, superscriptaddress]{revtex4-1}
\usepackage[T1]{fontenc}
\usepackage{color}
\usepackage{amsmath}
\usepackage{amssymb}
\usepackage{graphicx}
\usepackage{multirow}
\usepackage{booktabs}
\usepackage{capt-of}
\usepackage[latin9]{inputenc}
\usepackage{babel}
\usepackage{calc}
\usepackage{float}
\usepackage{dcolumn}

\makeatletter

\usepackage[colorlinks=true, letterpaper=true, pdfstartview=FitV, linkcolor=blue, citecolor=blue, urlcolor=blue]{hyperref}

\renewcommand{\fnum@figure}{Fig. \thefigure}
\renewcommand{\fnum@table}{Table \thetable}

\makeatother

\begin{document}

\title{Quantum metric-induced generalized magneto-optical effects in $\mathcal{PT}$-symmetric antiferromagnets}

\author{Yongpan Li}
\affiliation{Centre for Quantum Physics, Key Laboratory of Advanced Optoelectronic Quantum Architecture and Measurement (MOE), School of Physics, Beijing Institute of Technology, Beijing, 100081, China}

\author{Yichen Liu}
\affiliation{Centre for Quantum Physics, Key Laboratory of Advanced Optoelectronic Quantum Architecture and Measurement (MOE), School of Physics, Beijing Institute of Technology, Beijing, 100081, China}
 
\author{Cheng-Cheng Liu}
\email{ccliu@bit.edu.cn}
\affiliation{Centre for Quantum Physics, Key Laboratory of Advanced Optoelectronic Quantum Architecture and Measurement (MOE), School of Physics, Beijing Institute of Technology, Beijing, 100081, China}

\begin{abstract}
The magneto-optical effects (MOEs), as fundamental physical phenomena, can reveal the electronic structures of materials. The related probing methods are widely used in the study of magnetic materials. The conventional MOEs, i.e., Kerr and Faraday effects, are understood to arise from the Berry curvature (the imaginary part of the quantum geometry). Within the framework of conventional MOEs, space-time inversion ($\mathcal{PT}$) symmetric antiferromagnets are magneto-optically inactive. Here, we propose quantum metric (the real part of the quantum geometry) induced generalized MOEs, including the Voigt effect (or the quadratic Kerr effect) and more general magnetism-dependent polarization rotations, and build generic formulas with quantum metric for polarization rotation angles in three-dimensional and two-dimensional $\mathcal{PT}$-symmetric antiferromagnets. Combining the tight-binding model and first-principles calculations, we demonstrate the quantum metric-induced generalized MOEs in the $\mathcal{PT}$-symmetric antiferromagnets. Our theory reveals a novel quantum metric mechanism beyond the conventional Berry curvature paradigm and provides a microscopic understanding of the polarization rotations observed in $\mathcal{PT}$-symmetric antiferromagnets.
\end{abstract}

\maketitle

\textit{Introduction}---The magneto-optical effects (MOEs) have been a long-standing research focus~\cite{PhysRevLett.30.1329,PhysRevB.8.1239,KATO1995713,PhysRevB.51.12633,PhysRevLett.112.017205,feng_large_2015,huang2017layer,WOS:000524540200001,hu2023new,khaghani2022enhanced,PhysRevLett.131.156702,PhysRevB.107.014404,PhysRevB.109.094435,yin2021chiral,ahn2022theory,xu2022three,xia2022enhancement,kato2023topological,li2024topological,BCCD}. When light is incident on a magnetic material, both the reflection and transmission of the light at the surface, as well as the propagation of the light inside the material, can be characterized by the frequency-dependent complex refractive index. The fundamental physics underlying various MOEs is that the two eigenmodes of the magnetic material have different complex refractive indices. Given the material-specific dielectric tensor $\boldsymbol{\epsilon}$ and permeability tensor $\boldsymbol{\mu}$, we can obtain the complex refractive indices by solving the wave equation. Since one can take $\boldsymbol{\mu}=1$ at optical and near-infrared frequencies~\cite{landau_electrodynamics_1961,pershan1967magneto} and the optical conductivity tensor $\boldsymbol{\sigma}$ can be converted into the dielectric tensor $\boldsymbol{\epsilon}$~\cite{khj_buschow_handbook_1990,kuch_magnetic_2014}, the optical conductivity tensor $\boldsymbol{\sigma}$ of magnetic materials is a crucial physical quantity in the study of MOEs.

With the introduction of quantum geometry~\cite{peotta2015superfluidity, ahn_riemannian_2022,gao_field_2014,de2017quantized,torma2022superconductivity,PhysRevLett.132.026301,luhaizhou2024quantume,verma2024instantaneous,PhysRevX.14.011052,ezawa2024analytic,yu_quantum_2025,kim_direct_2025} in recent years, we have gained a more comprehensive understanding of the optical conductivity~\cite{JPSJ.12.570,PhysRevB.9.4897,sipe_nonlinear_1993,aversa_nonlinear_1995,sipe_secondorder_2000,jia_equivalence_2024}. The optical conductivity tensor $\boldsymbol{\sigma}$ can be decomposed into contributions from the quantum metric $\boldsymbol{\sigma}^g$ and Berry curvature $\boldsymbol{\sigma}^\Omega$. The conventional understanding holds that the vanishing Berry curvature-driven transverse optical conductivity, i.e., $\sigma_{xy}^\Omega=0$, would result in the disappearance of MOEs. A well-known example is the Kerr effect, where the Kerr angle is given by $\widetilde{\theta}_K=-\sigma_{xy}^{\Omega}/(\sigma_{xx}^g\sqrt{1+i\sigma_{xx}^g/\omega\epsilon_0})$~\cite{voigt1908magneto,PhysRev.186.891}. However, the significance of the quantum metric in the MOEs has not yet been appreciated.

\begin{table*}[t]
\caption{\label{tab:sym_ana}Symmetry-reduced form of the optical conductivity tensor $\boldsymbol{\sigma}$. Without loss of generality, we consider light incident along the positive $z$-direction on a surface lying in the $xy$-plane ($z=0$). NS denotes no symmetry constraint, $\mathcal{PT}$ refers to space-time inversion symmetry, $\mathcal{M}$ refers to mirror and mirror-like symmetry, and $\mathcal{O}$ refers to threefold or higher rotational symmetry. The contributions from the quantum metric and Berry curvature are denoted as $\sigma^g_{ij}$ and $\sigma^\Omega_{ij}$ according to Eq. (\ref{eq:oc_g_o}), respectively.}
\begin{ruledtabular}
\begin{tabular}{cccc}
NS&$\mathcal{PT}$&$\mathcal{M}$&$\mathcal{O}$\\
\colrule
$\begin{pmatrix}\sigma^g_{xx}&\sigma^g_{xy}+\sigma^\Omega_{xy}\\\sigma^g_{xy}-\sigma^\Omega_{xy}&\sigma^g_{yy}\end{pmatrix}$ & $\begin{pmatrix}\sigma^g_{xx}&\sigma^g_{xy}\\\sigma^g_{xy}&\sigma^g_{yy}\end{pmatrix}$ & $\begin{pmatrix}\sigma^g_{xx}&0\\0&\sigma^g_{yy}\end{pmatrix}$ & $\begin{pmatrix}\sigma^g_{xx}&\sigma^\Omega_{xy}\\-\sigma^\Omega_{xy}&\sigma^g_{xx}\end{pmatrix}$\\
\end{tabular}
\end{ruledtabular}
\end{table*}

In this Letter, we uncover a novel quantum metric mechanism for MOEs and propose the quantum metric-induced generalized MOEs in space-time inversion ($\mathcal{PT}$) symmetric antiferromagnets. The generalized MOEs describe magnetism-dependent optical polarization effects that include the Voigt effect~\cite{voigt1908magneto,PhysRevLett.87.047401,nyy3-h8wr,kimak2026ultrafast} (also known as the quadratic Kerr effect~\cite{10.1063/1.1449436,POSTAVA1997199,PhysRevB.68.064401,wohlrath2025quadratic,kimak2026ultrafast,PhysRevB.100.064403}) and more general magnetism-dependent polarization rotations. We develop the general formulas involving the quantum metric for generalized MOEs in three-dimensional (3D) and two-dimensional (2D) $\mathcal{PT}$-symmetric systems. Using the tight-binding model and first-principles calculations, based on our generic formulas, we show that the quantum metric can solely induce remarkable generalized MOEs. Our theory explains the polarization rotations observed in $\mathcal{PT}$-symmetric antiferromagnets, resolving discrepancies between existing theories and experiments, and challenges the notion that only Berry curvature drives MOEs.

\textit{Formulas for optical conductivity tensor}---The optical conductivity tensor $\boldsymbol{\sigma}$, which is defined by the constitutive relation $j_a=\sigma_{ab}E_b$, links the optical current density $\boldsymbol{j}$ to the optical electric field $\boldsymbol{E}$.  The optical conductivity tensor is a complex frequency-dependent quantity, characterizing the linear response of a material to an electromagnetic field. The explicit form of the components of the interband optical conductivity tensor $\boldsymbol{\sigma}$ is given by~\cite{aversa_nonlinear_1995}
\begin{align}
\sigma_{ab}(\omega)=&-\frac{ie^2}{\hbar}\sum_{n\neq m}\int\frac{d^3k}{(2\pi)^3}(f_{n\boldsymbol{k}}-f_{m\boldsymbol{k}})\nonumber\\
&\times\omega_{nm\boldsymbol{k}}\frac{r_{nm\boldsymbol{k}}^ar_{mn\boldsymbol{k}}^b}{\omega_{nm\boldsymbol{k}}+\omega+i\eta},
\end{align}
and the intraband optical conductivity tensor is given by~\cite{aversa_nonlinear_1995}
\begin{equation}\label{eq:intra_oct}
\sigma_{ab}^{\mathrm{Drude}}(\omega)=\frac{ie^2}{\omega+i\eta}\sum_{n}\int_{\boldsymbol{k}}\frac{\partial f_{n\boldsymbol{k}}}{\partial k_b}v^a_{nn\boldsymbol{k}}.
\end{equation}
$f_{n\boldsymbol{k}}$ is the Fermi-Dirac distribution function. $\hbar\omega_{nm\boldsymbol{k}}\equiv\hbar\omega_{n\boldsymbol{k}}-\hbar\omega_{m\boldsymbol{k}}$ is the energy difference between the $n$th band and $m$th band at the same $\boldsymbol{k}$ point. $\eta>0$ is a smearing parameter. $r_{nm\boldsymbol{k}}^a$ is the matrix element of the position operator. $v^a_{nn\boldsymbol{k}}\equiv\hbar\partial\omega_{n\boldsymbol{k}}/\partial k_a$.

Recent advances in quantum geometry~\cite{ahn_riemannian_2022, gao_field_2014, de2017quantized, torma2022superconductivity, PhysRevLett.132.026301, luhaizhou2024quantume, verma2024instantaneous, PhysRevX.14.011052, ezawa2024analytic} have provided a more advanced framework for understanding optical conductivity~\cite{JPSJ.12.570,PhysRevB.9.4897,sipe_nonlinear_1993,aversa_nonlinear_1995,jia_equivalence_2024,PhysRevB.109.115121}. Notably, the term $r_{nm\boldsymbol{k}}^ar_{mn\boldsymbol{k}}^b$ is identified as the components of the quantum geometry tensor $\boldsymbol{Q}$~\cite{provost_riemannian_1980,shapere1989geometric,ma_abelian_2010,ahn_riemannian_2022}, which can be rewritten as
\begin{equation}
Q^{ab}_{nm\boldsymbol{k}}:=r_{nm\boldsymbol{k}}^ar_{mn\boldsymbol{k}}^b\equiv g^{ab}_{nm\boldsymbol{k}}-\frac{i}{2}\Omega^{ab}_{nm\boldsymbol{k}},
\end{equation}
with $g^{ab}_{nm\boldsymbol{k}}$ and $\Omega^{ab}_{nm\boldsymbol{k}}$ being the components of the quantum metric tensor and Berry curvature tensor, respectively. Therefore, the transverse optical conductivity $\sigma_{xy}$ can be rewritten as [SM~\cite{SM} Sec. I]
\begin{align}\label{eq:oc_g_o}
\sigma_{xy}(\omega)=&\frac{2ie^2}{\hbar}\sum_{n\neq m}\int\frac{d^3k}{(2\pi)^3}\left(\frac{f_{n\boldsymbol{k}}\omega\omega_{mn\boldsymbol{k}}}{\omega_{mn\boldsymbol{k}}^2-(\omega+i\eta)^2}g_{nm\boldsymbol{k}}^{xy}\right.\nonumber\\
&\left.+\frac{f_{n\boldsymbol{k}}\omega_{mn\boldsymbol{k}}^2}{\omega_{mn\boldsymbol{k}}^2-(\omega+i\eta)^2}\frac{i}{2}\Omega_{nm\boldsymbol{k}}^{xy}\right)\nonumber\\
\equiv&\sigma_{xy}^g+\sigma_{xy}^\Omega,
\end{align}
where $\sigma_{xy}^g$ represents the contribution from the quantum metric $g_{nm\boldsymbol{k}}^{xy}$ and $\sigma_{xy}^\Omega$ represents the contribution from the Berry curvature $\Omega_{nm\boldsymbol{k}}^{xy}$. Note that since the diagonal components of the Berry curvature tensor are zero ($\Omega_{nm\boldsymbol{k}}^{xx}=\Omega_{nm\boldsymbol{k}}^{yy}=0$), the diagonal components of the optical conductivity tensor are exclusively contributed by the quantum metric, i.e., $\sigma_{xx}=\sigma_{xx}^g$ and $\sigma_{yy}=\sigma_{yy}^g$.

Since quantum geometry is a multi-band effect, the intraband optical conductivity is unrelated to quantum geometry~\cite{provost_riemannian_1980,ma_abelian_2010,peotta2015superfluidity,torma2022superconductivity,ahn_riemannian_2022}. Moreover, from Eq.~(\ref{eq:intra_oct}), it can be seen that the intraband optical conductivity is related to the properties of the Fermi surface of metals, while the materials in our first-principles calculations are insulators. Therefore, the intraband optical conductivity is irrelevant here.

\textit{Symmetry-reduced optical conductivity tensor}---We begin by studying the form of the optical conductivity tensor under symmetry constraints. Without loss of generality, consider light incident along the positive $z$-direction on a surface ($z=0$).

The Berry curvature $\Omega^{ab}_{nm\boldsymbol{k}}$ and quantum metric $g^{ab}_{nm\boldsymbol{k}}$ transform as $\Omega^{ab}_{nm\boldsymbol{k}}=-\Omega^{ab}_{nm\boldsymbol{k}}$ and $g^{ab}_{nm\boldsymbol{k}}=g^{ab}_{nm\boldsymbol{k}}$ under $\mathcal{PT}$-symmetry. Therefore, for $\mathcal{PT}$-symmetric antiferromagnets, the Berry curvature is zero everywhere in the Brillouin zone (BZ) and the optical conductivity tensor is purely contributed by the quantum metric.

The quantum metric $g^{xy}_{nm\boldsymbol{k}}$ and the Berry curvature $\Omega^{xy}_{nm\boldsymbol{k}}$ are symmetric and antisymmetric under exchange of $x$ and $y$, respectively, i.e., $g^{xy}_{nm\boldsymbol{k}}=g^{yx}_{nm\boldsymbol{k}}$ and $\Omega^{xy}_{nm\boldsymbol{k}}=-\Omega^{yx}_{nm\boldsymbol{k}}$. $\sigma_{xy}^g$ and $\sigma_{xy}^\Omega$ obey the same symmetry restrictions as the quantum metric and Berry curvature, i.e., $\sigma_{xy}^g=\sigma_{yx}^g$ and $\sigma_{xy}^\Omega=-\sigma_{yx}^\Omega$. However, threefold or higher rotational symmetry imposes constraints on the form of the optical conductivity tensor~\cite{kleiner_spacetime_1966}, i.e., $\sigma_{xx}^{g}=\sigma_{yy}^{g}$ and $\sigma_{xy}^{g/\Omega}=-\sigma_{yx}^{g/\Omega}$. Therefore, in this case, the quantum metric makes no contribution to the transverse optical conductivities $\sigma_{xy}$ and $\sigma_{yx}$.

Another category of symmetries that one should pay attention to is mirror and mirror-like symmetry $\mathcal{M}$. The mirror symmetries $M_x$ or $M_y$ can forbid the transverse optical conductivities, i.e., $\sigma_{xy}^g=\sigma_{yx}^g=0$. We collectively refer to other symmetries that impose the same restrictions ($\sigma_{xy}^g=\sigma_{yx}^g=0$) on the optical conductivity tensor as mirror-like symmetries, including glide symmetries ($M_x\tau$ and $M_y\tau$), two-fold rotational symmetries ($C_{2x}$ and $C_{2y}$), and two-fold screw symmetries ($C_{2x}\tau$ and $C_{2y}\tau$).

The symmetry analysis above is summarized in Table \ref{tab:sym_ana}, and the detailed derivations can be found in Supplementary Materials (SM)~\cite{SM} Sec. I.

\textit{Theory of quantum metric-induced generalized MOEs}---The wave equations in $\mathcal{PT}$-symmetric antiferromagnets are given by [SM~\cite{SM} Sec. II]
\begin{equation}\label{eq:sys_of_eq_eigenmode}
\begin{pmatrix}-n^2+1+i\sigma_{xx}^g/\omega\epsilon_0&i\sigma_{xy}^g/\omega\epsilon_0\\i\sigma_{xy}^g/\omega\epsilon_0&-n^2+1+i\sigma_{yy}^g/\omega\epsilon_0\end{pmatrix}\begin{pmatrix}E_x\\E_y\end{pmatrix}=0,
\end{equation}
where $n$ is the complex refractive index and $\epsilon_0$ is the vacuum permittivity. The complex refractive indices $n_\pm$ of the eigenmodes of light in a material can be obtained by solving the wave equations, i.e.,
\begin{equation}\label{eq:complex_refractive_for_lr}
n_\pm=\sqrt{1-\frac{\sigma_{xx}^g+\sigma_{yy}^g\pm \sqrt{(\sigma_{yy}^g-\sigma_{xx}^g)^2+4(\sigma_{xy}^g)^2}}{2i\omega\epsilon_0}}.
\end{equation}
Our formulas are universal since we do not assume high symmetry, like cubic symmetry in conventional MOE theory~\cite{khj_buschow_handbook_1990,kuch_magnetic_2014}, as well as $\sigma_{xx}\gg\sigma_{xy}$ in typical cases~\cite{khj_buschow_handbook_1990,kuch_magnetic_2014} and $\sigma_{xx}\ll\sigma_{xy}$ in topological magnetic thin films~\cite{tse_giant_2010,tse_magnetooptical_2011}.

For $\mathcal{PT}$-symmetric antiferromagnets, the polarization rotation angle of the reflected light, $\widetilde{\theta}_R=\theta_R+i\xi_R$, and the polarization rotation angle of the transmitted light
$\widetilde{\theta}_T=\theta_T+i\xi_T$, are given by [SM~\cite{SM} Sec. II]
\begin{align}
\widetilde{\theta}_R&=\frac{2\sigma_{xy}^g}{\sigma_{xx}^g-\sigma_{yy}^g+\sqrt{(\sigma_{yy}^g-\sigma_{xx}^g)^2+4(\sigma_{xy}^g)^2}R},\label{eq3k}\\
\widetilde{\theta}_T&=\frac{2\sigma_{xy}^g}{\sigma_{xx}^g-\sigma_{yy}^g+\sqrt{(\sigma_{yy}^g-\sigma_{xx}^g)^2+4(\sigma_{xy}^g)^2}T},\label{eqfara}
\end{align}
where $R$ and $T$ exhibit different forms in 3D and 2D systems, capturing the distinctions between the reflection and transmission coefficients of the two eigenmodes, respectively. For 3D systems, $R$ and $T$ are given by
\begin{equation}
R^{\mathrm{3D}}=\frac{n_+n_--1}{n_+-n_-},\ T^{\mathrm{3D}}=\frac{e^{ik_0n_+l}+e^{ik_0n_-l}}{e^{ik_0n_+l}-e^{ik_0n_-l}},
\end{equation}
where $l$ is the propagation distance of the light. For 2D systems, $R$ and $T$ read
\begin{equation}
R^{\mathrm{2D}}=\frac{r(n_+)+r(n_-)}{r(n_+)-r(n_-)},\ T^{\mathrm{2D}}=\frac{t(n_+)+t(n_-)}{t(n_+)-t(n_-)},
\end{equation}
$t(n_i)$ and $r(n_i)$ are the total complex transmission coefficient and the total complex reflection coefficient, respectively, and depend on the shape of the sample and substrate. In SM~\cite{SM} Sec. III, we provide expressions for $t(n_i)$ and $r(n_i)$ in a relatively simpler case. When the system exhibits threefold or higher rotational symmetry, our general framework for the 3D systems can recover the commonly known forms of Kerr and Faraday angles~\cite{voigt1908magneto,PhysRev.186.891}, and a detailed discussion is given in SM~\cite{SM} Sec. II.

Equations (\ref{eq:complex_refractive_for_lr}-\ref{eqfara}) represent the main result of this work and capture the physical significance of the quantum metric in MOEs. In materials with low-symmetry lattices, Eqs. (\ref{eq3k}) and (\ref{eqfara}) may contain a nonmagnetic contribution from lattice optical anisotropy. To identify Eqs. (\ref{eq3k}) and (\ref{eqfara}) unambiguously as generalized MOEs, we focus on systems whose nonmagnetic reference structures possess threefold or higher rotational symmetry, so that the nonmagnetic background is symmetry-forbidden.

\begin{figure}[t]
\centering
\includegraphics[width=1.0\linewidth]{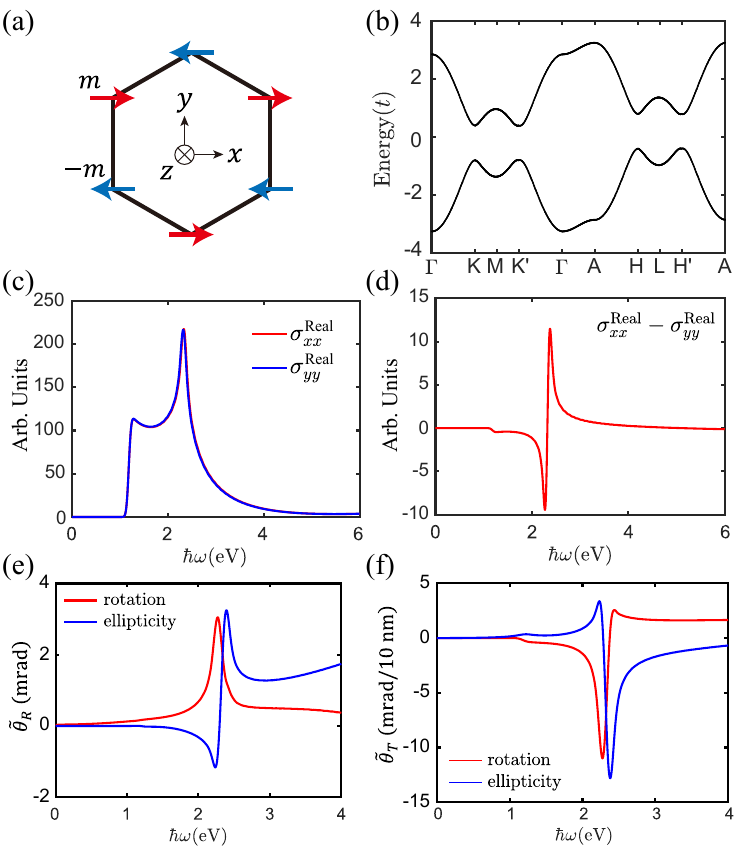}
\caption{The quantum metric-induced generalized MOEs in $\mathcal{PT}$-symmetric antiferromagnets. (a) Schematic of the tight-binding model. The red and blue arrows denote up and down spins, respectively. (b) The spin-degenerate energy bands. (c) The real part of the diagonal terms of the optical conductivity tensor, $\sigma_{xx}$ and $\sigma_{yy}$. As explained in the text, the imaginary part can be obtained from the corresponding real part through the Kramers-Kronig relation. (d) The difference between the real parts of the diagonal terms of the optical conductivity tensor. (e) The reflection polarization rotation angle $\widetilde{\theta}_R=\theta_R+i\xi_R$ is shown in terms of the polarization rotation $\theta_R$ and polarization ellipticity $\xi_R$. (f) The transmission polarization rotation angle $\widetilde{\theta}_T/l=\theta_T/l+i\xi_T/l$ is shown in terms of the polarization rotation $\theta_T$ and polarization ellipticity $\xi_T$ per unit thickness $l$, with $l$=10 nm. In (e) and (f), the optical electric field direction is along the $\boldsymbol{x}+\boldsymbol{y}$ direction. Parameters: $m=0.6t$, $t_{\mathrm{SOC}}=t_z=0.1t$, and the smearing parameter is set to $0.04t$.}
\label{fig:tbmodel}
\end{figure}

\textit{Tight-binding model}---We now consider a 3D honeycomb lattice with collinear $\mathcal{PT}$-symmetric antiferromagnetic order as shown in Fig.~\ref{fig:tbmodel}(a). The Hamiltonian is given by~\cite{liu_lowenergy_2011}
\begin{align}
H=&-\sum_{\langle ij\rangle,\alpha}tc_{i\alpha}^\dagger c_{j\alpha}+\sum_{i\alpha\beta}\xi_imc_{i\alpha}^\dagger s^x_{\alpha\beta}c_{i\beta}\nonumber\\
&-t_z\sum_{i\alpha}\left(c_{i\alpha}^\dagger c_{i+z,\alpha}+c_{i\alpha}^\dagger c_{i-z,\alpha}\right)\nonumber\\
&+it_{\text{SOC}}\sum_{\langle\!\langle ij\rangle\!\rangle,\alpha\beta}\xi_{i}c_{i\alpha}^\dagger[(\boldsymbol{s}\times\hat{\boldsymbol{d}}_{ij})_z]_{\alpha\beta} c_{j\beta}.
\end{align}
Here, $c_{i\alpha}^\dagger$ ($c_{i\alpha}$) creates (annihilates) an electron at site $i$ with spin $\alpha$, and $\xi_i = \pm1$ takes opposite signs for the two sublattices. The first term is the intralayer nearest-neighbor hopping term, while the third term is the interlayer vertical hopping term. The second term denotes the antiferromagnetic order with a N\'eel vector aligned along the $x$ direction. The final term incorporates an intralayer next-nearest-neighbor intrinsic Rashba spin-orbit coupling (SOC). $[(\boldsymbol{s}\times\hat{\boldsymbol{d}}_{ij})_z]_{\alpha\beta}=s^x_{\alpha\beta}d_{ij}^y-s^y_{\alpha\beta}d_{ij}^x$, where $\hat{\boldsymbol{d}}_{ij}$ represents the unit vector pointing from site $j$ to $i$.

In the absence of magnetic order ($m=0$), the system possesses threefold rotation symmetry $C_{3z}$. Consequently, potential background signals that could interfere with MOEs, such as the nonmagnetic optical anisotropy, are strictly forbidden, since $C_{3z}$ suppresses the contribution from the quantum metric and $\mathcal{PT}$ prohibits the contribution from the Berry curvature. We introduce an $x$-directed N\'eel order to break the $C_{3z}$ symmetry while maintaining the $\mathcal{PT}$ symmetry. Since the lattice exhibits no nonmagnetic optical anisotropy, the generalized MOEs are driven entirely by the magnetic order, i.e., the generalized MOEs vanish when $m=0$ and emerge only when $m\neq0$.

With the antiferromagnetic order, the system still has $\mathcal{PT}$-symmetry, leading to spin degeneracy in all bands as shown in Fig.~\ref{fig:tbmodel}(b). In Fig.~\ref{fig:tbmodel}(c), the real parts of the diagonal terms ($\sigma_{xx}$ and $\sigma_{yy}$) of the optical conductivity tensor are shown, and the off-diagonal terms ($\sigma_{xy}$ and $\sigma_{yx}$) of the optical conductivity tensor are zero due to the mirror symmetry $M_y$. Since the optical conductivity tensor is contributed purely by the quantum metric, the real part and the imaginary part of the optical conductivity tensor are just the Hermitian part and anti-Hermitian part of the optical conductivity tensor, respectively [SM~\cite{SM} Sec.~V]. Therefore, the imaginary part can be obtained from the corresponding real part through the Kramers-Kronig relation. The difference between $\sigma_{xx}$ and $\sigma_{yy}$ is shown in Fig.~\ref{fig:tbmodel}(d), which explicitly shows the breaking of threefold rotational symmetry by the antiferromagnetic order.

The reflection and transmission polarization rotation angles are shown in Figs. \ref{fig:tbmodel}(e) and \ref{fig:tbmodel}(f), respectively. When the energy of the photon is smaller than the minimum value of the local gap, there is no transmission polarization rotation and reflection polarization ellipticity, while reflection polarization rotation and transmission polarization ellipticity are finite. These behaviors in quantum metric-induced generalized MOEs stand in contrast to those in conventional MOEs, where Kerr rotation and Faraday ellipticity are absent, but Faraday rotation and Kerr ellipticity are finite~\cite{khj_buschow_handbook_1990}.

\textit{Materials.}---Using the first-principles method, we apply our generalized MOE theory to $\mathcal{PT}$-symmetric antiferromagnets. Here, we take two illustrative $\mathcal{PT}$-symmetric antiferromagnetic materials, i.e., 2D monolayer $\mathrm{MnPSe_3}$~\cite{WIEDENMANN19811067}, 3D bulk $\mathrm{CaMn_2Sb_2}$~\cite{BRIDGES20093653}, as examples. Although both 2D monolayer $\mathrm{MnPSe_3}$ and 3D bulk $\mathrm{CaMn_2Sb_2}$ crystallize in structures with threefold rotational symmetry, their antiferromagnetic orders break the threefold rotational symmetry.

\begin{figure}[t]
\centering
\includegraphics[width=1.0\linewidth]{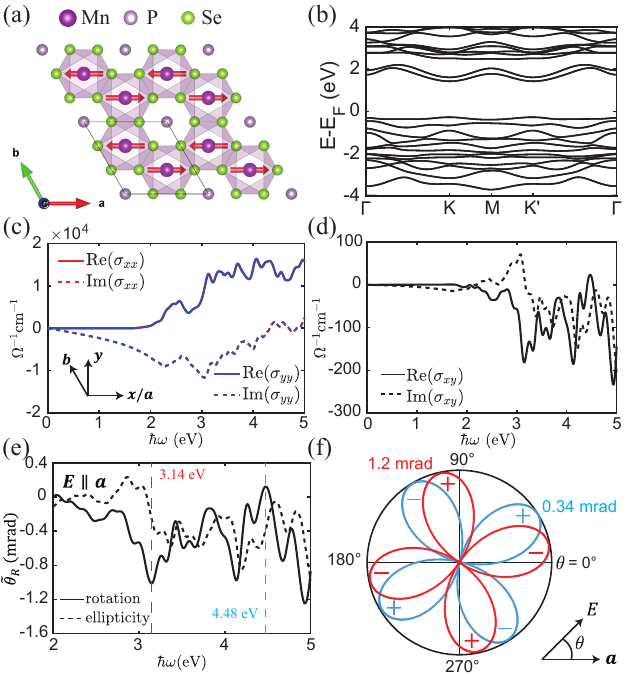}
\caption{The quantum metric-induced generalized MOEs in 2D $\mathcal{PT}$-symmetric antiferromagnets. (a) Crystal structure of monolayer $\mathrm{MnPSe_3}$. The crystal structure is drawn using VESTA~\cite{vesta}. (b) The band structure of monolayer $\mathrm{MnPSe_3}$. (c) The longitudinal conductivities $\sigma_{xx}$ and $\sigma_{yy}$, with the $x$- and $y$-axes defined in the inset. (d) The transverse conductivities $\sigma_{xy}=\sigma_{yx}$. (e) The reflection polarization rotation angle spectrum for the optical electric field polarized along the $a$-axis. (f) Angular dependence of the reflection polarization rotation with respect to the optical electric field direction, plotted at selected photon energies. The red and blue curves correspond to photon energies of 3.14 eV and 4.48 eV, respectively. The radial coordinate represents the magnitude of the polarization rotation, and the azimuthal coordinate represents the angle between the electric field and the $a$-axis. The blue or red $+$ and $-$ signs indicate the signs of the corresponding polarization rotation.}
\label{fig:MnPSe3}
\end{figure}

\begin{figure}[t]
\centering
\includegraphics[width=1.0\linewidth]{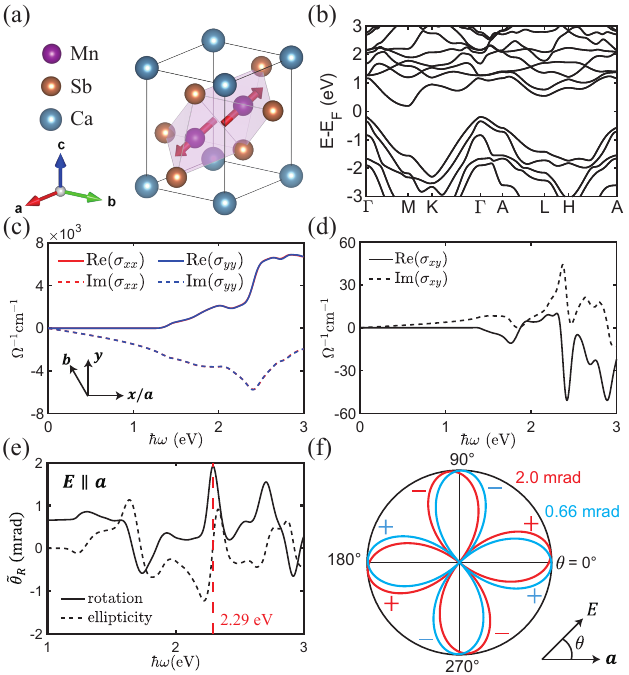}
\caption{The quantum metric-induced generalized MOEs in 3D $\mathcal{PT}$-symmetric antiferromagnets. (a) Crystal structure of bulk $\mathrm{CaMn_2Sb_2}$. (b) The band structure of bulk $\mathrm{CaMn_2Sb_2}$. (c) The longitudinal conductivities $\sigma_{xx}$ and $\sigma_{yy}$. The inset defines the $x$- and $y$-axes, where the $x$-axis coincides with the $a$-axis. (d) The transverse conductivities $\sigma_{xy}=\sigma_{yx}$. (e) The photon energy dependence of the reflection polarization rotation. The inset shows the optical electric field direction, aligned with the $a$-axis. (f) The dependence of the reflection polarization rotation on the direction of the optical electric field for specific photon energies. The blue and red curves correspond to photon energies of 1.00 eV and 2.29 eV, respectively. The radial coordinate represents the magnitude of the reflection polarization rotation, and the azimuthal coordinate represents the angle between the electric field and the $a$-axis. The blue or red $+$ and $-$ signs indicate the signs of the corresponding reflection polarization rotation.}
\label{fig:CaMn2Sb2}
\end{figure}

Both 2D monolayer $\mathrm{MnPSe_3}$ and 3D bulk $\mathrm{CaMn_2Sb_2}$ are $\mathcal{PT}$-symmetric antiferromagnets with BNS magnetic space group $P\overline{1}'$ (No. 2.6). In the nonmagnetic phase, 2D monolayer $\mathrm{MnPSe_3}$ and 3D bulk $\mathrm{CaMn_2Sb_2}$ belong to space group $R\overline{3}$ (No. 148) and $P\overline{3}m1$ (No. 164), respectively. The crystal structures and magnetic orders of $\mathrm{MnPSe_3}$ and $\mathrm{CaMn_2Sb_2}$ are shown in Fig.~\ref{fig:MnPSe3}(a) and Fig.~\ref{fig:CaMn2Sb2}(a). Owing to the threefold rotational symmetry preserved in both crystal structures, no nonmagnetic polarization rotation emerges to confuse the generalized MOEs, rendering Eqs. (\ref{eq3k}) and (\ref{eqfara}) identically zero in the absence of magnetic order. Therefore, a non-zero polarization rotation in $\mathrm{MnPSe_3}$ and $\mathrm{CaMn_2Sb_2}$ serves as a definitive signature of the generalized MOEs.

The band structures in Fig.~\ref{fig:MnPSe3}(b) and Fig.~\ref{fig:CaMn2Sb2}(b) show that the monolayer $\mathrm{MnPSe_3}$ and bulk $\mathrm{CaMn_2Sb_2}$ are insulators. The optical conductivity tensor for $\mathrm{MnPSe_3}$ is given in Fig.~\ref{fig:MnPSe3}(c) and Fig.~\ref{fig:MnPSe3}(d), and the optical conductivity tensor for $\mathrm{CaMn_2Sb_2}$ is presented in Fig.~\ref{fig:CaMn2Sb2}(c) and Fig.~\ref{fig:CaMn2Sb2}(d). We study the reflection polarization rotation angle as a function of photon energy and optical electric field orientation. The reflection polarization rotation angle spectra with the optical electric field along the $a$-axis are shown in Fig.~\ref{fig:MnPSe3}(e) for $\mathrm{MnPSe_3}$ and Fig.~\ref{fig:CaMn2Sb2}(e) for $\mathrm{CaMn_2Sb_2}$. When the optical electric field is oriented along the $a$-axis, $\mathrm{MnPSe_3}$ exhibits the maximum reflection polarization rotation of approximately 1.2 mrad at 4.93 eV, while $\mathrm{CaMn_2Sb_2}$ shows a remarkable reflection polarization rotation of about 1.9 mrad at 2.29 eV. We also investigate the dependence of the reflection polarization rotation on the direction of the optical electric field. As shown in Fig.~\ref{fig:MnPSe3}(f) and Fig.~\ref{fig:CaMn2Sb2}(f), the reflection polarization rotation exhibits anisotropic characteristics, and the optical electric field direction that maximizes the reflection polarization rotation varies with photon energy.

To demonstrate the magnetic origin of the generalized MOEs without ambiguity, we have focused on materials whose nonmagnetic parent phases possess threefold or higher rotational symmetry, such that nonmagnetic optical anisotropy is symmetry-forbidden. Nevertheless, $\mathcal{PT}$-symmetric antiferromagnets with low-symmetry lattices constitute an experimentally important class of materials~\cite{xu2021Observation, zhang_cavityenhanced_2022, guo_colossal_2024,gish_van_2024}, where the quantum metric-induced generalized MOEs coexist with nonmagnetic optical anisotropy. Experimentally, the generalized MOEs can be extracted by comparing the optical responses measured in the magnetic and nonmagnetic states, which is a standard experimental strategy~\cite{PhysRevLett.116.097204, nanolett.1c01719,6gbq-d7jk}. We also analyze several representative $\mathcal{PT}$-symmetric antiferromagnets with low-symmetry lattices, including strained bilayer $\mathrm{CrI_3}$, monolayer $\mathrm{FePS_3}$, monolayer and bulk $\mathrm{CoPS_3}$, and bulk $\mathrm{CoAgPO_4}$, in SM~\cite{SM} Sec.~IV.

\textit{Discussion.}---Previously, the role of the quantum metric in MOEs had not been recognized, and the symmetry of MOEs was regarded as equivalent to the symmetry of Berry curvature. Our work demonstrates that in the $\mathcal{PT}$-symmetric antiferromagnets, one must consider the contribution from the quantum metric, which leads to the generalized MOEs. It should be noted that even for 3D $\mathcal{PT}$-symmetric antiferromagnets with threefold or higher rotational symmetry, the quantum metric-induced generalized MOEs remain active when light is incident on their low-symmetry surfaces. Therefore, many materials previously thought to lack MOEs, such as $\mathcal{PT}$-symmetric antiferromagnets, actually exhibit generalized MOEs. Our work reveals purely quantum metric-induced generalized MOEs in $\mathcal{PT}$-symmetric antiferromagnets and greatly deepens the understanding of MOEs.

While the $\mathcal{PT}$ symmetric antiferromagnets with vanishing Berry curvature are magneto-optically inactive within the framework of conventional MOEs, recent experimental works~\cite{saidl2017optical, yang2019magneto, little2020three,xu2021Observation, zhang_cavityenhanced_2022, guo_colossal_2024,gish_van_2024} have revealed giant reflection polarization rotation in $\mathcal{PT}$-symmetric antiferromagnets, and our generalized MOE theory provides a reasonable explanation. In particular, our calculated reflection polarization rotations show agreement with experimental measurements~\cite{xu2021Observation}.

\textit{Acknowledgements.}---We acknowledge support from the NSF of China (Grant No. 12374055), the Science Fund for Creative Research Groups of NSFC (Grant No. 12321004), and the National Key R\&D Program of China (Grant No. 2020YFA0308800).

\bibliography{reference}
\end{document}